\documentclass[authoryear]{elsarticle}
\usepackage{graphicx}
\usepackage{amsmath,amsfonts}
\usepackage{subfigure}
\usepackage{array, multirow}
\usepackage{endfloat} % Move floats to the end of the document
\usepackage{multirow}

\usepackage[pdftex, usenames]{color}
\definecolor{darkblue}{rgb}{0,0,0.5}
\definecolor{darkgreen}{rgb}{0,0.5,0}
\usepackage[pdftex, colorlinks]{hyperref}
\DeclareMathOperator*{\argmax}{arg\,max}

\newif\ifhavefigures
\havefigurestrue

\journal{Evolution} 
\begin{document}
\begin{frontmatter}
%\title{Phylogenetic Comparative Methods Require Statistically Robust Model Comparisons}
\title{Is your phylogeny informative? \\Measuring the power of comparative methods}
\author[cpb]{Carl Boettiger\corref{cor1}}
\ead{cboettig@ucdavis.edu}
\author[cpb]{Graham Coop}
\author[cpb]{Peter Ralph}
%\cortext[cor1]{Corresponding author.}
\address[cpb]{Center for Population Biology, University of California, Davis, United States}
\address[eve]{Department of Evolution and Ecology, University of California, Davis, United States}
\begin{abstract}
Phylogenetic comparative methods may fail to produce meaningful results when either the underlying model is inappropriate
or the data contain insufficient information to inform the inference.  
The ability to measure the statistical power of these methods has become crucial 
to ensure that data quantity keeps pace with growing model complexity.
Through simulations, we show that commonly applied model choice methods based on information criteria 
can have remarkably high error rates;
this can be a problem because methods to estimate the uncertainty or power are not widely known or applied.
Furthermore, the power of comparative methods can depend significantly on the structure of the data.
We describe a Monte Carlo based method which addresses both of these challenges,
and show how this approach both quantifies and substantially reduces errors relative to information criteria.
The method also produces meaningful confidence intervals for model parameters.
We illustrate how the power to distinguish different models, 
such as varying levels of selection, 
varies both with number of taxa and structure of the phylogeny.  
We provide an open-source implementation in the \texttt{pmc} (``Phylogenetic Monte Carlo'') 
package for the \texttt{R} programming language.
We hope such power analysis becomes a routine part of model comparison in comparative methods.  
\end{abstract}
	
\end{frontmatter}

\section{Introduction}
\subsection{Are phylogenies informative?}
% Paragraph Summary: Phylogenetic methods are ubiquitous
Since their introduction into the comparative method over two and a half decades ago,
phylogenetic methods have become increasingly common and increasingly complex.
Despite this, concern persists about the ubiquitous use of these approaches~\citep{Price1997,Losos2011c}.  
From a statistical perspective these concerns can be divided into two categories: 
(a) Do we have appropriate models that reflect the biological reality of evolution and represent meaningful hypotheses, and 
(b) Do we have adequate data to fit these models and to choose between them?
The models have been greatly improved since their introduction, 
and can now account for stabilizing selection~\citep{Hansen1996}, multiple optima~\citep{Butler2004}, 
and differing rates of evolution across taxa~\citep{O'Meara2006} 
or through time~\citep{Pagel1999, Blomberg2003};
but little attention has been given to this second concern about data adequacy.
In this paper, we highlight the importance of these concerns, and illustrate a method for addressing them.

% Paragraph Summary: the need for power estimates
It can be difficult to accurately interpret the results of comparative methods
without quantification of uncertainty, model fit, or power. 
% Guess this is a problem of ML only when it isn't bootstrapped or equivalent
% Guess it's not a problem for Bayes per se, 
Most current comparative methods do not attempt to quantify this uncertainty;
consequently it can be easy for inadequate power to lead to false biological conclusions.
For instance, below 
% should I drop the ``we/show we illustrate'' part?
we illustrate how estimates of phylogenetic signal~\citep{Gittleman1990} using the $\lambda$ statistic~\citep{Pagel1999, Revell2010} 
can reach opposite conclusions (from no signal $\lambda = 0$ to approximately Brownian, $\lambda \approx 1$) 
when applied to different simulated realizations of the same process.
We also show that model selection by information criteria can prefer over-parameterized models by a wide margin.  
On the other hand, when a simpler model is chosen,
it may be difficult to determine whether this merely reflects a lack of power.  
In both cases, the results can be correctly interpreted by estimating the uncertainty in parameter estimates
and the statistical power (ability to distinguish between models) of the model selection procedure.

%Paragraph Summary: We draw attention to this & provide a method to address it
Here we provide one solution to these problems using a parametric bootstrapping approach 
which easily fits within the framework used by many comparative methods approaches. 
As comparative methods rely on explicit models, this is easily implemented by simulating under the specified models.
For the problem of uncertainty in parameter estimation, the bootstrap is a well-established and straightforward method~\citep{Efron1987}.  
A few areas of comparative methods have used a similar approach:
for instance, phylogenetic ANOVA~\citep{Garland1993} 
calculates $p$ values of the test statistic by simulation under Brownian motion.  
A similar approach was later introduced in the \texttt{Brownie} software~\citep{O'Meara2006}
to generate the null distribution of likelihood ratios under Brownian motion, and applied in
\citet{Revell2008a}, which showed the distribution can deviate substantially from $\chi^2$.  
Unfortunately, such approaches have never become a common in comparative analyses.  
Here, we describe a method due to \citet{Cox1962} and used by others~\citep{Goldman1993,Huelsenbeck1996},
that can be used in place of information criteria for model choice,
allowing estimation of power and false positive rates,
and can provide good estimates of confidence intervals on model parameter estimates.  
While simulations are often performed when a new method is first presented, this practice rarely becomes routine.  
By providing a simple R package (``\texttt{pmc}'', phylogenetic Monte Carlo) for the method outlined, 
we hope Monte Carlo based model choice and estimates of power become common in comparative methods.  

To set the stage, we will review common phylogenetic models 
and describe the Monte Carlo approach to model choice.
We then present the results of our method applied to example data and discuss its consequences.

\subsection{Common phylogenetic models} \label{ss:models}
Comparative phylogenetics of continuous traits commonly uses 
a collection of simple stochastic models of evolution;
we briefly review these here to fix ideas and notation.
All models we consider take as given an ultrametric phylogenetic tree whose branch lengths represent evolutionary divergence times;
extant taxa are represented by the tips of the tree.
We will assume that the tree is known without error. 
For convenience we will in all examples choose time units so that the tree height is one unit.
%(While it is possible to relax this assumption it is not yet typical of comparative phylogenetic analyses.)  
For each extant taxon we have a trait value (say, the species mean) for some continuous trait such as body size, and represent the collection of trait values across extant taxa as the vector $X$.  
%(It is possible to consider a suite of traits for each species simultaneously, though almost all analyses focuses on a single trait at a time.)
The joint distribution of these trait values is given by specifying the ancestral trait value $X_0$ at the root of the tree, 
by describing the stochastic process of trait evolution along branches of the tree,
and assuming that evolution on separate branches proceeds independently.

Let $Y_t$ be the value of our trait at time $t$ along some branch.  
The simplest and most common model for the evolution of the trait $Y_t$ is
a scaled Brownian motion~\citep{Felsenstein1985}, 
which can be represented by the stochastic differential equation:
\begin{equation}
dY_t = \sigma dB_t ,
\label{bm}
\end{equation}
in which $B_t$ is standard Brownian motion, and $\sigma$ is the rate parameter.
Under this model, 
the trait value evolves as a random walk starting from the ancestral state $X_0$,
and upon reaching each node in the phylogeny, the process bifurcates into two independent Brownian walks.  
This Brownian motion (BM) model is completely defined given a phylogeny and two parameters: 
the initial state $X_0$ and the parameter $\sigma$, which is usually interpreted as the rate of increase in variance.
%or simply, the ``trait diversification rate.''  

A closely related model introduced in a comparative phylogenetics context by \citet{Hansen1997} 
is the Ornstein-Uhlenbeck (OU) model, for which 
trait evolution $Y_t$ along each branch follows the Ornstein-Uhlenbeck process, 
which is described by the following stochastic differential equation
\begin{equation}
dY_t = - \alpha (Y_t - \theta)dt + \sigma d B_t .
\label{ou}
\end{equation}
Here Brownian motion is modified to have a central tendency towards a preferred trait value $\theta$, 
usually interpreted as a optimum trait value under stabilizing selection.  
The strength of stabilizing selection increases linearly with distance from the optimum $\theta$, 
controlled by the parameter $\alpha$.  
When $\alpha = 0$, this model reduces to the BM model.  
Both evolutionary models are described in more detail elsewhere, \emph{e.g.}~\citet{Butler2004}.

Many variations of these basic models are also common -- for instance, it may be desirable to allow the diversification rate parameter $\sigma$ in the BM model to vary in some way over time~\citep{Blomberg2003, Harmon2010, Pagel1999} or across the phylogeny \citep{O'Meara2006}.  
Similar extensions can be applied to the OU model -- we will later consider the example of \citet{Butler2004} which allows the optimum trait value $\theta$ to differ among different branches or clades.
One can illustrate which branches of a phylogeny are permitted to have independently estimated values of the optimum trait by ``painting'' them different colors indicating where the model is allowed to change~\citep{Butler2004}.  

Another commonly used variation is Pagel's $\lambda$~\citep{Pagel1994, Freckleton2002}, 
which was introduced as a test of phylogenetic signal
-- the degree to which correlations in traits reflect patterns of shared ancestry.  
The model underlying Pagel's $\lambda$ is the simple Brownian motion along the phylogeny as above,
except that the phylogeny is modified by shortening all internal edges by a multiplicative factor of $\lambda$,
which reduces the resulting correlations between any pair of taxa by a factor $\lambda$, 
and adjusting terminal edges so the tree remains ultrametric.
The parameter $\lambda$ can then be estimated by maximum likelihood.
Estimates near unity are taken to indicate high phylogenetic signal,
% (this is not how it is interpreted?) indicating that evolution appears Brownian.  
while estimates near zero 
indicate that other processes such as natural selection have erased this ``signal'' of common descent.

\section{Methods} %% Methods

\subsection{Uncertainty in parameter estimates}
To demonstrate the perils of inadequate data without estimates of uncertainty,
we open with an example of a phylogenetic test using Pagel's $\lambda$ statistic
that also serves to illustrate the estimation of uncertainty in parameter estimates
(e.g.\ confidence intervals).
We illustrate that on a small tree,
estimates of $\lambda$ can differ greatly from the parameter used in the simulations.
In practice, the danger is that an estimate of $\lambda$ near zero may arise by chance because the tree is too small,
not because the phylogeny is unimportant to the evolution of the trait. 
Larger phylogenies, on the other hand, generally allow greater accuracy.

\begin{figure}[hht]
\begin{center}
\ifhavefigures
    \subfigure[ ]{\includegraphics[width=2.2in]{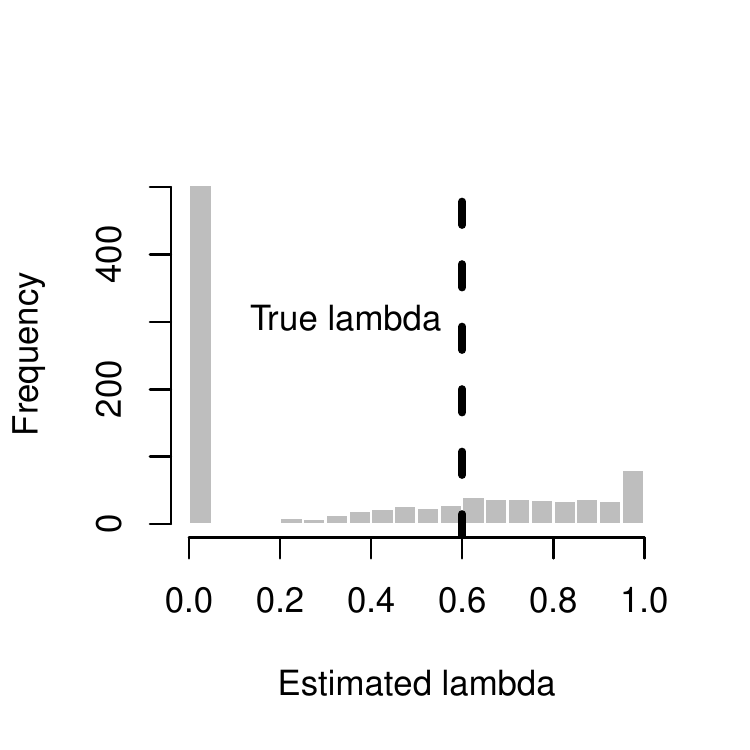}\label{fig:geospiza_lambda}}
    \subfigure[ ]{\includegraphics[width=2.2in]{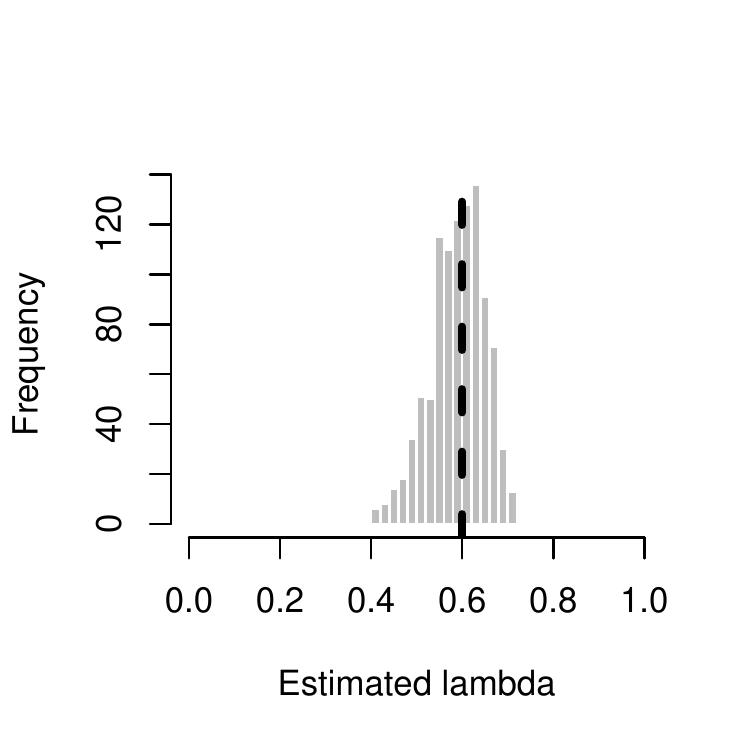}\label{fig:simtree_lambda}}
    \subfigure[ ]{\includegraphics[width=2.2in]{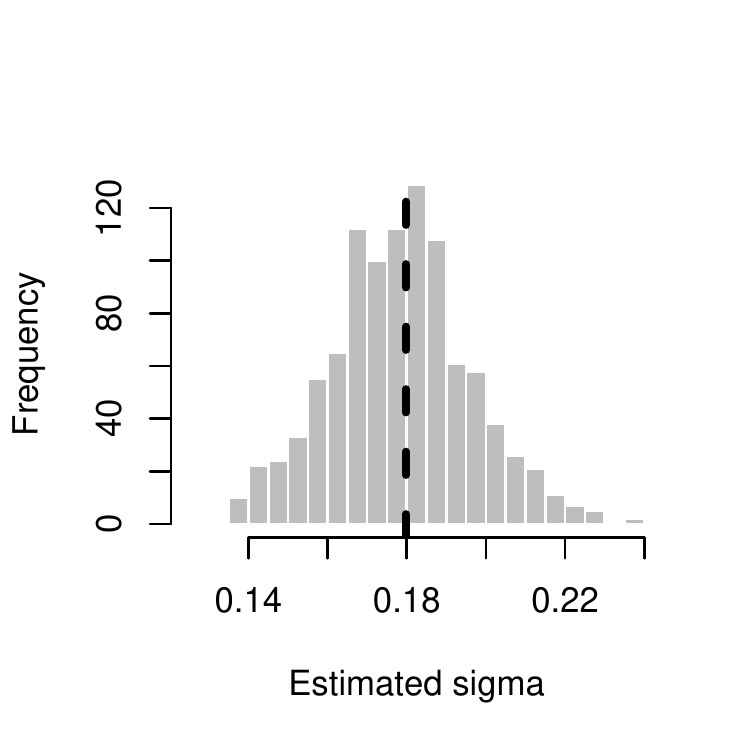}\label{fig:geospiza_sigma}}
\fi
\end{center}
\caption{(a) Empirical distribution of maximum likelihood estimates of $\lambda$ 
for 1000 sets of trait values simulated on the {\em Geospiza} phylogeny with 13 taxa
transformed with $\lambda=0.6$, using $\sigma=0.18$.
Most such datasets yielded a maximum likelihood estimate of 0;
the mean estimate is $\bar \lambda = 0.35$.  
% Estimates fail to recover the true value used in the simulation even when averaged over replicates.  
% Phylogenetic tree of 13 taxa of \emph{Geospiza} finches included in the \texttt{geiger} package. 
(b)  As above, but simulating trait values on a much larger phylogenetic tree 
(a single, simulated Yule tree with 281 tips), again transformed with $\lambda=0.6$.
The estimated values now cluster around the true value, 
and have mean $\bar \lambda = 0.59$.  
(c) The data can be more informative about some parameters than others: 
shown is the empirical distribution of maximum likelihood estimates of the diversification rate $\sigma$ for the same simulations as in (a).
The mean of the distribution is $\bar \sigma = 0.18$, matching the value used in the simulations.  
}
\label{fig:lambda}
\end{figure}

In Figure~\ref{fig:geospiza_lambda} we show the empirical distribution
of the maximum likelihood estimate of $\lambda$
for 1000 data sets simulated under 
a model with moderate phylogenetic signal, $\lambda =0.6$,
and $\sigma=.03$.
The estimates were performed on the \emph{Geospiza} data using functions available in {\texttt pmc} in conjunction with the R package \texttt{geiger}~\citep{Harmon2008}.
The phylogeny, data, and script for the analysis are included in {\texttt pmc}.
We see that for datasets coming from this small phylogeny, 
the maximum likelihood statistic $\hat \lambda$
is a poor estimator for the true value of $\lambda$. 
The most common estimate is $\hat \lambda = 0$,
which is usually interpreted to mean that the phylogeny contains little information.  
The next most common estimate is $\hat \lambda = 1$. 
Note that this is the upper bound set on $\lambda$ by the fitting algorithm.
% Even averaging across all datasets,
% we consistently underestimate $E(\hat \lambda) = 0.4$ the true value of parameter used to simulate the data, ($\lambda =0.6$).  % not needed.
It is clear that 
we must thus be cautious what we conclude based on values of $\lambda$ estimated on this phylogeny.
% This danger is not unique to estimating phylogenetic signal,
% but arises whenever we attempt to draw conclusions based on the estimate of a model parameter.  

Repeating this exercise on successively larger data sets makes it clear that this is a problem of insufficient data.  
With a simulated tree of 281 tips,
the estimated values are closely centered around the true value, as shown in Figure~\ref{fig:simtree_lambda}.

The amount of data required to be informative will depend
not only on the size and topology of the tree but also on the question being asked.  
For instance, it may be impossible to distinguish moderately different values of $\lambda$, 
which is very difficult to estimate accurately.
However, it may be feasible to estimate other parameters on smaller phylogenies than this 281 taxa example.
For instance, 
using the same 13 taxa {\it Geospiza} phylogeny, 
we can estimate the diversification rate parameter $\sigma$ much more precisely, 
as shown in Figure~\ref{fig:geospiza_sigma}.

A natural way to report the uncertainty associated with a parameter estimate 
is construct a confidence interval,
which is rarely performed in the literature but
can easily be done by parametric bootstrapping.
Given the parameter estimate, a confidence interval can be estimated by 
simulating a large number of datasets using the known phylogeny and the estimated parameter, 
and re-estimating the parameter on each simulated dataset~\citep[\emph{e.g.} see][]{Diciccio1996}.
The distribution of the re-estimated parameters is used to construct the confidence interval;
e.g.\ the $2.5$ to the $97.5$ percentile gives a $95\%$ confidence interval.
For the example shown in Figure~\ref{fig:simtree_lambda}, our estimate of $\lambda$ on the Yule tree with 281 tips,
the $95\%$ confidence interval would be $(0.45, 0.69)$.  
For the parameter $\sigma$, Figure~\ref{fig:geospiza_sigma} shows that the confidence interval is $(0.007, 0.059)$.
Given the noisy nature of parameters estimated from phylogenies we recommend that confidence interval should routinely be reported,
and to facilitate this, have implemented this as \texttt{pmc::confidenceIntervals.pow}.
Confidence intervals could also be estimated from the curvature of the likelihood surface, 
but this is often infeasible or inaccurate.  
For instance,
the interval for $\lambda$ on the 281 tip tree computed using the Hessian is $ \pm 73.1$,
clearly not a good approximation of the interval computed directly from the simulation, $\pm 0.12$, Figure~\ref{fig:simtree_lambda}.

\begin{figure}[hht]
\begin{center}
\includegraphics[width=12cm]{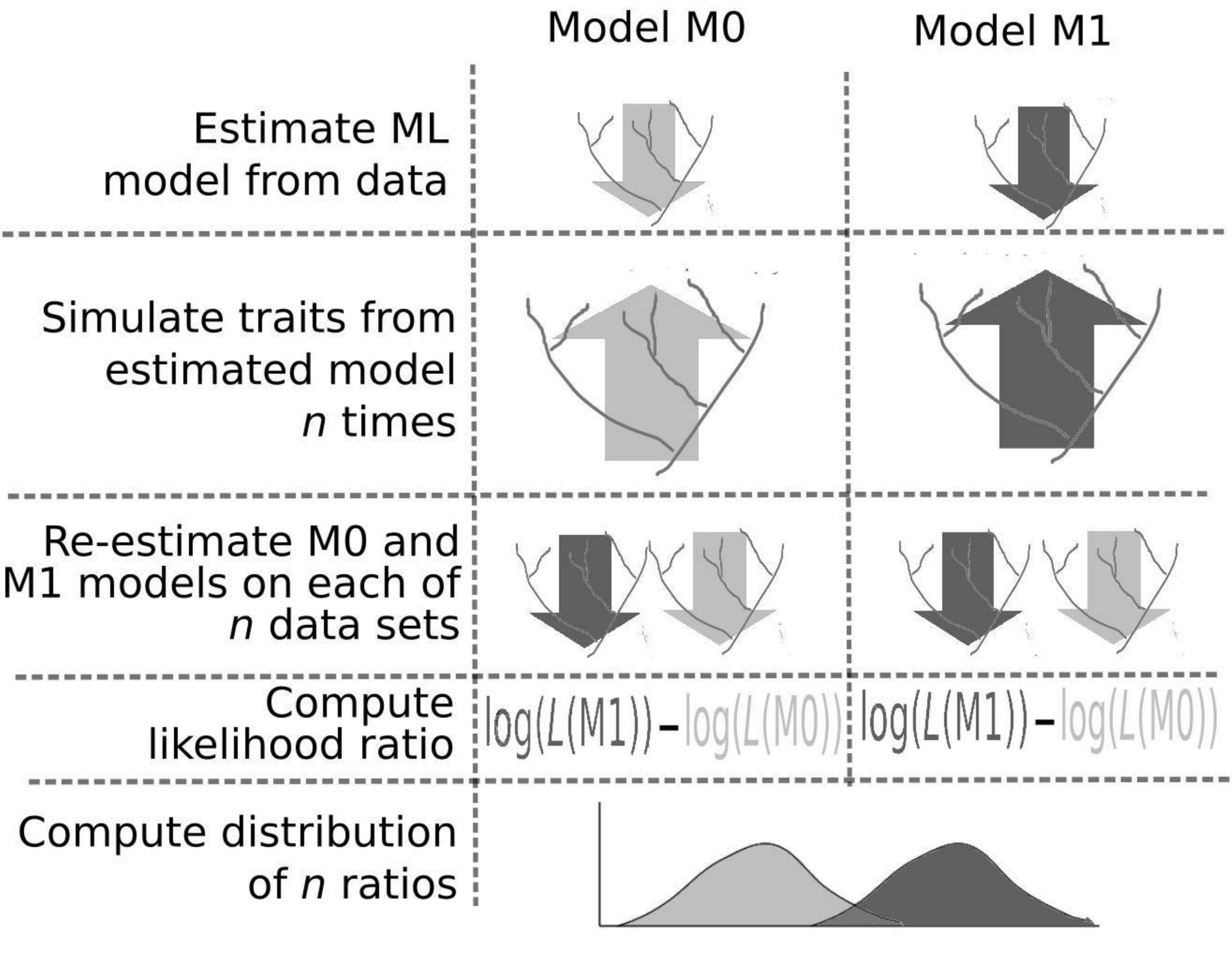}\label{fig:concept} 
\end{center}
\caption{
%(a) 
Conceptual diagram of the Monte Carlo method for model choice.  
First, parameters for both models are estimated from the original data. 
Then, $n$ simulated datasets are created from each model at these parameters, 
and on each dataset,
the parameters for both models are re-estimated
and the likelihood ratio statistic is computed.
The collection of likelihood ratio statistics generates the corresponding distribution.  
This involves a process of $4n$ fits by maximum likelihood, instead of only 2 fits required for information criteria.
%(b) Visualization of $p$ value in the distribution of the likelihood ratio statistic.  Integrating the distribution of $\delta$ generated by simulating under $M_0$ to the right of the observed value in the original data, ~\citep{Cox1961,Cox1962}.  (c) Visualization of the power associated with this test: the integral of the distribution of $\delta$ generated by simulating under $M_1$, from the right of the 5\% tail of the distribution under model $M_0$.
% Not sure the p/power figures are sufficiently valuable
}
\end{figure}

\subsection{The Monte Carlo approach}

Knowing when the data are sufficiently informative is also crucial when comparing different models.  
% Model comparison has become a routine part of phylogenetic analysis. 
% As models increase in complexity to test more subtle features of the evolutionary process,
% it becomes essential to quantify how informative the data actually are of a particular inference.  
To do this, we introduce a Monte Carlo-based method, described below.
Suppose we have a dataset $X$ for which we wish to determine which of two models, model 0 or model 1, is the better description.  
Each model is specified by a vector of parameters, $\Theta_0$ and $\Theta_1$ respectively, 
which can assume values in the spaces $\Omega_0$ and $\Omega_1$ respectively.  
We tend to imagine that model 1 is the more complex model,
though in general they need not be nested.
Let $\mathcal{L}_0$ be the likelihood function for model 0, 
let $\hat \Theta_0 = \argmax_{\Theta_0 \in \Omega_0}(\mathcal{L}_0(\Theta_0|X))$ 
be the maximum likelihood estimator for $\Theta_0$ given $X$, 
and let $L_0 = \mathcal{L}_0 (\hat \Theta_0 | X)$; 
and define $\mathcal{L}_1$, $\hat \Theta_1$, $L_1$ similarly for model 1.

The statistic we will use is $\delta$, 
defined to be twice the difference in log likelihood of observing the data under the two MLE models, % (following~\citet{Huelsenbeck1996})
\begin{equation}
\delta = - 2\left( \log L_0 - \log L_1 \right)
\label{delta}
\end{equation}
%The Neyman-Pearson Lemma~\citep{Neyman1933} demonstrates that this is the most powerful comparison between models.
For simplicity we will refer to this as the likelihood ratio. 
Larger values of $\delta$ indicate more support for model 1 relative to model 0. 
% If the models are nested than $\delta$ is always positive; otherwise this may not hold.  
It is natural to use the difference in log-likelihoods as a statistic to choose between the models~\citep{Neyman1933},
as do information criteria such as AIC.
To do this we need to know, for instance, how large should $\delta$ be before we decide that model 1 is much closer to the truth than is model 0.
Many common methods proceed to approximate the distribution of $\delta$ asymptotically.
For instance, if the models are nested in a manner that does not force a parameter to its boundary value,
this statistic has asymptotically the $\chi^2$ distribution with degrees of freedom equal to the difference in the number of parameters.
%(This is one justification for the AIC criterion.)
These asymptotic approximations for phylogenetic comparative analyses are often inadequate for phylogenetic comparisons.  
Instead, we can estimate the distribution of $\delta$ under either model directly from Monte Carlo simulation.  
This method seems to have been first suggested in the statistical literature by \citet{Cox1961, Cox1962} 
and applied to mixture models by \citet{McLachlan1987}.
It has been previously applied to the case of estimating phylogenies from sequence data by \citet{Huelsenbeck1996};
see also \citet{Goldman1993}.  

To estimate the distribution of $\delta$ under model 0 and the estimated parameters ($\hat \Theta_0$), we proceed as follows.
First simulate $n$ datasets $X^1, \ldots, X^n$ independently from model 0 with parameters $\hat \Theta_0$.  
For each $1 \le k \le n$, let $\hat \Theta_0^k$ be the maximum likelihood estimator of the parameters $\Theta_0$ of model 0 for dataset $X^k$, 
and likewise let $\hat \Theta_1^k$ be the MLE under model 1.
Then we compute the likelihood ratio statistic for the $k^{\text{th}}$ data set, $\delta_k = - 2 ( \log \mathcal{L}_0( X^k | \hat \Theta^k_0 ) - \log \mathcal{L}_1( X^k | \hat \Theta^k_1 ) )$,
and examine the empirical distribution of $\delta_1, \ldots, \delta_n$.
We can also estimate the distribution of $\delta$ under model 1 in the same way.

There are two things to note about this procedure. 
First, the Monte Carlo datasets are simulated at the maximum likelihood parameters $\hat \Theta_0$ and $\hat \Theta_1$, 
which are in turn estimated from the same dataset $X$.  
So if, for instance, the models are nested and the simpler is correct, 
then one would expect model 0 at $\hat \Theta_0$ to be quite similar to model 1 at $\hat \Theta_1$.
%One may be tempted to treat this as a comparison of ``model 0 versus model 1'',
%across all parameter values $\Theta_0$ and $\Theta_1$, but this is not quite the case.
Secondly, it is necessary when computing the Monte Carlo values $\delta_k$ to {\em re-estimate} the maximum likelihood parameters,
rather than using the original parameters $\hat \Theta_0$ and $\hat \Theta_1$ --
simply computing $\delta_k = - 2 ( \log \mathcal{L}_0(X^k | \hat \Theta_0 ) - \log \mathcal{L}_0(X^k | \hat \Theta_1 ) )$
would lead to a much less powerful test~\citep{Hall1991a}.
The reason for this is somewhat subtle \citep[see][]{McLachlan1987}, and is related to the first point.  
For further suggestions on obtaining a reliable estimate of the distributions, 
see \citet{Efron1987} and \citet{Diciccio1996}.

\subsection{Model selection}
If we suppose model 0 is ``simpler'' than model 1, it is natural to regard model 0 as the ``null'' and test the hypothesis that the data came from model 0.
To do this, we would compare where the observed difference in log likelihoods $\delta$ for the original data falls relative to the distribution under model 0.
The proportion of the simulated values larger than $\delta$ 
%(see Figure~\ref{fig:shade_p}) 
provides an approximation to the $p$--value for the test, the probability that a difference at least as large 
would be seen under model 0.
(Because the datasets $X^k$ are all simulated at the estimated parameters $\hat \Theta_0$ this strictly applies only for the hypothesis test between the maximum-likelihood estimated models, and is not the $p$-value when comparing the composite hypothesis represented by the original model with unspecified parameters~\citep[see][]{McLachlan1987}.)
If we choose, say, $\delta_*$ so that 95\% of the simulated values $\delta_1, \ldots, \delta_n$ fall below $\delta_*$,
and choose to reject model 0 if $\delta > \delta_*$, then we have a test of the null hypothesis that model 0 is true, with a false positive probability of approximately $.05$ under model 0.
% We say ``model 0'', but we should remember that the model in question is really ``model 0 at $\hat \Theta_0$'' --
% to compare model 0 to model 1 we would need to e.g.\ place priors on their respective parameter spaces.
If we then want to know about the statistical power of this test --
the probability that we correctly reject model 0 when the data came from model 1 --
we would turn to the distribution of $\delta$ under model 1.
% Broadly speaking, the amount of overlap between the distributions tells us the power the test has to distinguish between models.  
If we have chosen $\delta_*$ as above,
then the amount of this distribution to the left of $\delta_*$ approximates the probability of rejecting model 0 when the data are produced by model 1
-- the power of the test. 
%-- shown in Figure~\ref{fig:shade_power}.   

The procedure we have described is motivated by classical hypothesis testing,
but is only one way to use the information provided by the empirical distributions of $\delta$.

\section{An example using {\it Anolis} data}\label{sect:Anoles}
\subsection{The anoles data}
To illustrate the concerns about phylogenetic information in comparative methods, we shall revisit a classic data set of mean body size for 23 species of {\it Anolis} lizards from the Lesser Antilles, which has been used to introduce other comparative phylogenetic approaches~\citep[e.g.][familiar to many who have used the \texttt{ouch} package]{Butler2004}.  
The phylogeny reconstruction used here~\citep{Losos1990} is based upon  morphological \citep{Lazell1972} and protein-electrophretic \citep{Gorman1976} techniques rather than the more recent phylogenies based on mitochondrial sequences \citep{Schneider2001, Stenson2004}, which have substantial differences.  As our purpose is simply to illustrate the approach, we continue to use older tree familiar to the readers of earlier work~\citep{Losos1990, Butler2004}.

\begin{figure}[hht]
\begin{center}
\ifhavefigures
    \includegraphics[width=.8\textwidth]{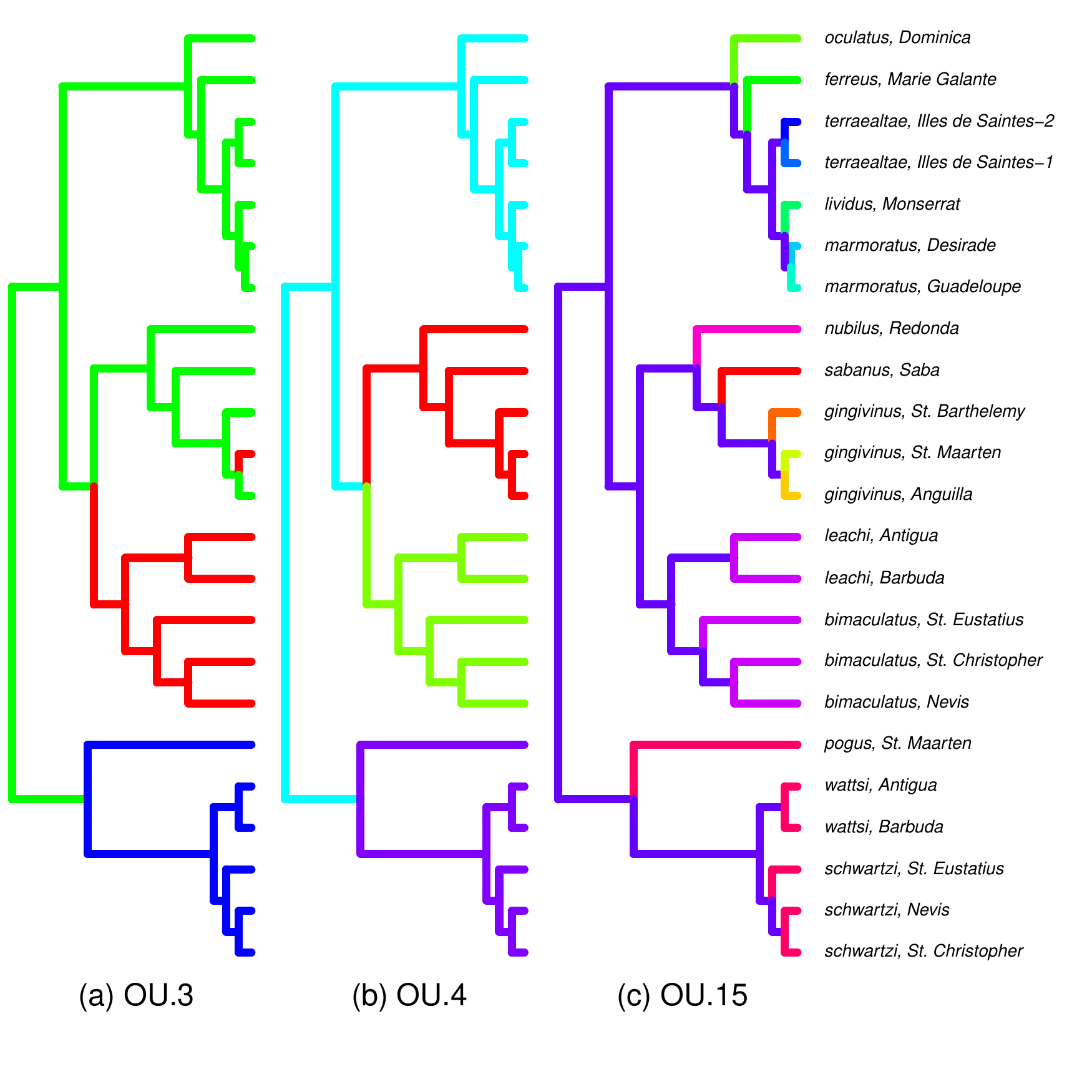}
\fi
\end{center}
\caption{
``Paintings'' of the {\it Anolis} phylogeny specifying which branches are assumed to have a common value of the trait optimum $\theta$ for three different models:
(a) {\bf OU.3}, with three possible optima from \citet{Butler2004};
(b) {\bf OU.4}, with four possible optima; and
(c) {\bf OU.15}, with a unique optimum for each branch in the upper two clades.
%\plr{ only 12 of these colors are distinguishable. Replace with letters?}
The remaining models, BM and OU.1, fit the same parameters across the entire phylogeny and so are not shown.  
Estimated model parameters for each are shown in Table~\ref{tab:pars}.}

\label{fig:paint}
\end{figure}

Identification of branches or clades of a phylogenetic tree that show significantly different evolutionary patterns can illuminate key elements about the origin and maintenance of biodiversity. 
\citet{Butler2004} demonstrated how the existence of different adaptive optima in character traits on different parts of a phylogenetic tree could be detected.  
They assumed that evolution of the trait along each branch followed the Ornstein-Uhlenbeck model, 
but that different branches could have different optima (the parameter $\theta$). 
The branches that must share a common value of $\theta$ are represented by a ``painting'' of the tree;
three possibilities for the {\it Anolis} tree that we later investigate are shown in Figure~\ref{fig:paint}.  
Any branch of a given color must have the same optimum trait value, each of which is estimated by the fitting algorithm.  
The remaining parameters $\alpha$ and $\sigma$ are shared across the entire tree.

To confirm that the proposed pattern of heterogeneity (the painting) is justified by the data, 
it is necessary to compare between possible paintings and possible assignments of model parameters 
to each part of the painting.
We seek to identify (a) which model best describes the data and 
(b) whether we have sufficient data to resolve that difference?

\subsection{Models for the {\it Anolis} phylogeny}
To illustrate the approach we consider a total of five models for the {\it Anolis} data set.  
The first two models apply the same model of evolution to the entire tree (i.e.\ a one-color painting) -- 
either Brownian motion (BM)~\citep{Edwards1964,Felsenstein1985}, with two parameters; 
or the Ornstein-Uhlenbeck process (OU.1)~\citep{Felsenstein1985,Hansen1997}, with three.
%\footnote{To completely specify the model, one also needs a root state.  This is taken as the maximum likelihood estimate, which is its expected value under BM model (often called the phylogenetic mean).}  

\begin{table}
  \begin{center}
  \begin{tabular}{|l|l|p{4.2cm}|p{7.8cm}|}
%\begin{tiny}
\hline
Model & $\log(L)$ & \footnotesize MLE Parameters & 95\% CI 
%& Description 
\\
\hline
BM &
17.33 &
\footnotesize $X_0=2.9$, 

$\sigma^2 = 0.043$ 
&
\footnotesize

(0.14, 0.26) 

(2.74, 3.16)
\\
\hline

OU.1 & 
15.69 &
\footnotesize $\theta=3.0 $, 

$\sigma^2 =0.048 $, 

$\alpha = 0.19$  
%& A single peak Ornstein-Uhlenbeck model.~\citep{Hansen1996} 
& % CI 
\footnotesize
(2.36, 3.56)

(0.028, .13)

(.24, 4.41)
\\
\hline

OU.3 &
24.82 & 
\footnotesize $\theta=\{3.36, 3.04, 2.56 \}$,

\footnotesize $\sigma^2 =0.05 $,

$\alpha =2.61 $ 
% & The most parsimonious reconstruction of colonization, \citep{Butler2004}, Fig~\ref{fig:paint}(a) 
& % CI
\footnotesize
$\{(3.20, 3.47 ), (2.94,3.11), (2.41,2.76 ) \}$

(0.025, 0.19)

(1.77, 17.98 )
\\
\hline
OU.4 & 
26.69 & 
\footnotesize $\theta=\{2.97, 3.31, 3.12, 2.63 \}$,

\footnotesize $\sigma^2 =0.06 $, 

$\alpha =4.68 $ 
% & A more complicated model with 4 optima corresponding to different clades, Fig~\ref{fig:paint}(b)  
& %CI
\footnotesize
$\{(2.87, 3.05 ), (3.22, 3.38 ), (3.02, 3.21 ), (2.53, 2.74) \}$

(0.031, 3.39)

(3.34, 384.16)
\\
\hline
OU.15 &
44.17 &
%\begin{center}
\footnotesize $\theta=\{2.91, 2.99, 2.98, 3.04,$ 
$3.11, 3.35, 2.97, 3.08,$
$3.19, 3.15, 3.17, 2.81,$
$ 3.30, 3.05, 2.62 \}$, 

\footnotesize  $\sigma^2 =0.06 $, 

$\alpha =24.3 $
%\end{center}
% &  An arbitrary, complex model using 15 different optima, Fig~\ref{fig:paint}(c) 
&  %CI 
\footnotesize
$\{$ (2.84,2.98 ), (2.91,3.22 ), (2.81,3.46 ), (2.85,3.57 ) 

\{(3.04,3.20 ), (3.28,3.53 ), (2.80,3.42 ), (2.30 , 352 ),

(2.34, 352 ), (2.94, 3.84 ) (2.94,3.85 ), (2.66, 1.5e6 ),

(3.27,3.38 ), (2.98, 3.12 ), (2.55, 2.67 ) $\}$

(0.0036, 0.44)

(7.29, 322.92)
\\
\hline
%\end{tiny}
\end{tabular}
\label{tab:pars}
\caption{Parameter values estimated for the {\it Anolis} dataset by maximum likelihood 
for models with varying number and location of optima (also see Figure~\ref{fig:paint}), used in the comparisons in Figure~\ref{fig:pmc}. 
These parameter values were used to produce the simulated datasets in Figures~\ref{fig:pmc} and \ref{fig:aic_errors}.
% Note that the comparisons consider maximum-likelihood estimated values of the models, consequently the models are defined not only by the paintings of Figure~\ref{fig:paint} but by the specific values of the parameters given.  
The values of $\theta$ are in order of first appearance, left-to-right, in Figure~\ref{fig:paint}.
The corresponding 95\% confidence intervals calculated from the 2000 replicates are also shown.  
Note that the optima $\theta$ of OU.15 just represent a finer partition of the optima in OU.3.}
\end{center}
\end{table}

The remaining three models extend these simple cases by introducing heterogeneity in the model, 
allowing the trait optimum to vary across the tree as indicated in Fig~\ref{fig:paint}.  
The OU.3 model of Figure~\ref{fig:paint}(a) has three optima, and corresponds to the {\em character displacement hypothesis}~\citep{Losos1990}, which predicts three different optimum body sizes -- an intermediate optimum on islands having only one species, and a larger and a smaller optimum for islands with two species of lizards.  
The island size determines to which optimum the tips or extant species are assigned, while the ancestral states are constructed by parsimony as per~\citet{Butler2004}.  
To these three models (BM, OU.1, and OU.3) analyzed by~\citet{Butler2004} we add two more to illustrate possible outcomes. 
OU.4, Figure~\ref{fig:paint}(b) hypothesizes four optima corresponding to four separate clades.  
The fifth model OU.15 is intentionally arbitrary and overly complex, assigning a unique optimum to each species in the top two clades for a total of 15 optima.   
We apply these methods to determine which model best fits the data and whether the data are sufficiently informative to distinguish between them.   

\begin{figure}[hht]
\begin{center}
\ifhavefigures
\includegraphics[width=\textwidth]{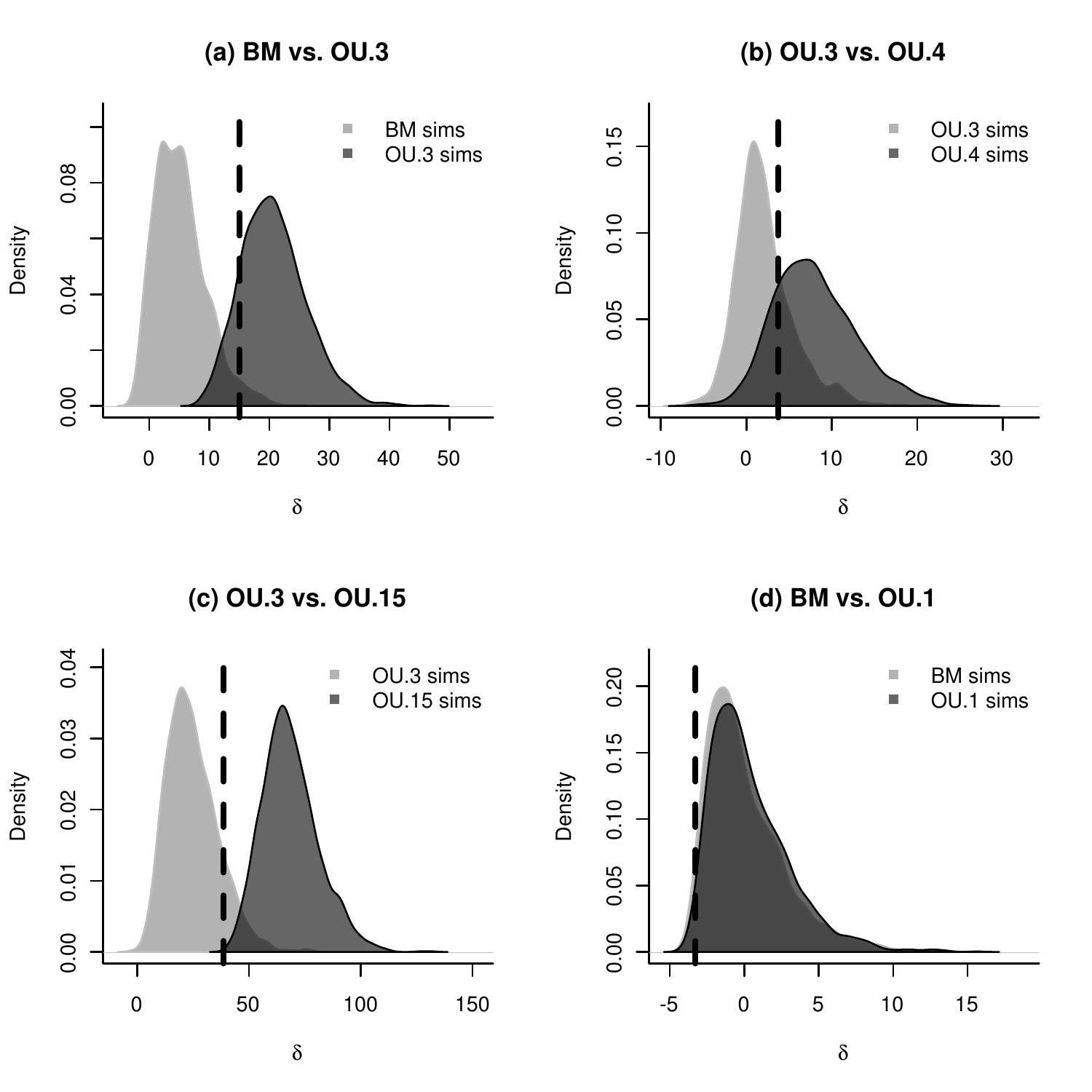}
\fi
\end{center}
\caption{
Distributions of the likelihood ratio statistic of equation \eqref{delta} for four different model comparisons.
In each case the lighter distribution shows the distribution of $\delta$ values 
obtained by bootstrapping under the simpler of the two models, 
while the darker distribution shows the distribution under the more complicated of the two models.
2000 replicates are used for each distribution.
The dashed vertical line indicates the observed value of $\delta$ when the models are fit to the {\it Anolis} dataset.
{\bf (a) BM versus OU.3:} the observed likelihood ratio is much more likely under OU.3.
{\bf (b) OU.3 versus OU.4:} here the distributions overlap more, indicating that the data are less informative about this more subtle comparison.  
{\bf (c) OU.3 versus OU.15:} these distributions have little overlap and the observed ratio falls clearly in the range of the simpler model. 
We can conclude that this support for OU.3 is not merely due to lack of power.  
{\bf (d) BM versus OU.1:} the data contain almost no information to distinguish between these two models at the estimated (small) level of selection $\alpha$. 
}
\label{fig:pmc}  
\end{figure}

\section{Results}

We illustrate several points with four different comparisons, depicted in Figure \ref{fig:pmc}.  
In each case, the distribution of $\delta$ under each of the two models is shown as the dark-shaded and the light-shaded curves,
and the observed value of $\delta$ is marked by the dashed vertical line. 
We also construct confidence intervals for the parameters in the same way as we did for the $\lambda$ estimates, shown in Table~\ref{tab:pars}.
The maximum likelihood parameter values for each model, estimated from the anoles data, are given in Table~\ref{tab:pars}, 
and are computed from the original body size data described in Section~\ref{sect:Anoles} using the \texttt{ouch} package of~\citet{Butler2004}
together with tools from our \texttt{pmc} package.  
Scripts to perform all analyses shown here are included in the \texttt{pmc} package.  
We will be able to determine not only which model is preferred, 
but also the certainty of the model choice.
% still not totally happy with this

\subsection{Quantification of model choice}

For a first example, comparing BM to OU.3 (Figure~\ref{fig:pmc}(a)),
we see that only 2.5\% of simulations under BM have a likelihood ratio $\delta$ more extreme
than the observed ratio of 15 units seen in the real data (\emph{i.e.} $p=0.025$). 
The degree of overlap in the distributions reflects 
the extent to which the phylogeny is useful to discriminate between the two hypotheses 
at these parameter values;
in this case the test that rejects the BM model with 5\% false positive rate has a power of 93.6\%. 
Thus we have a direct estimate of both which model is a better fit and of our power to choose between the models.  
Note that in our framework we are free to choose the tradeoff between the false positive and false negative rates.  
For instance, a 5\% cutoff may be too stringent if it is unnatural to treat either model as a null.   

\subsection{Information criteria often fail to choose the correct model}

For a second example, we compare OU.3 to the over-parameterized model OU.15 (Figure~\ref{fig:pmc}(c)).
Table \ref{tab:pars} shows that the maximum likelihood optimum trait values $\theta$ and rate of divergence $\sigma$ are similar for the two models,
but that the strength of selection $\alpha$ is much larger for OU.15.
From the table of estimated values and confidence intervals, 
it is clear that OU.15 has simply divided up each of these broader peaks into finer optima clustered around the original estimates.  
The higher value of $\alpha$ in the OU.15 model indicates narrow peaks of strong selection that result in the much higher likelihood.  
Despite this, our method will not select OU.15,
since the observed likelihood ratio $\delta$ falls below value of $\delta$ seen in 18.8\% of simulations under OU.3.
Furthermore, this is a powerful test:
98.8\% of simulations under OU.15 produce a $\delta$ that falls beyond the 95\% quantile of the OU.3 distribution. 

\begin{figure}[hht]
\begin{center}
    \includegraphics[width=5in]{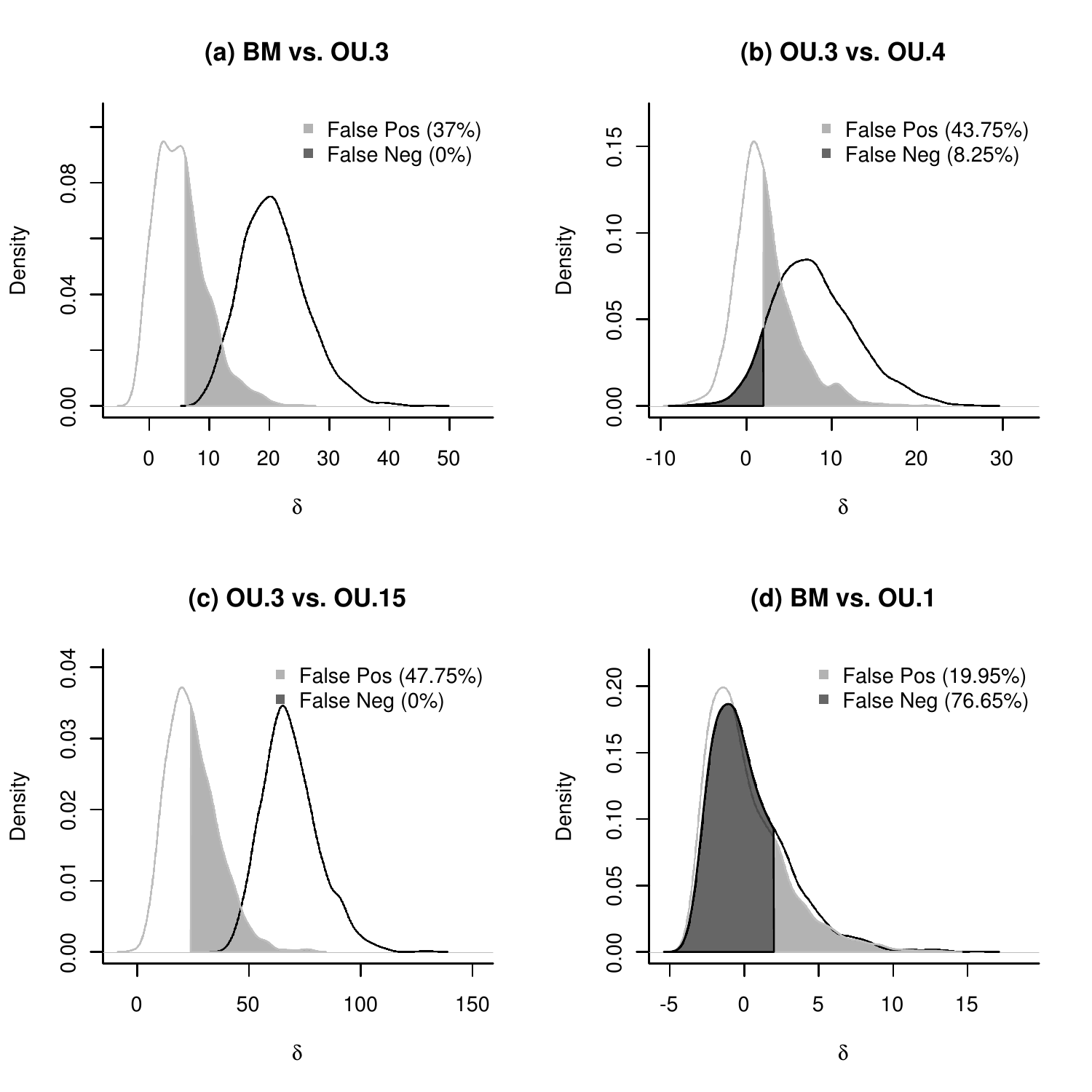}
\end{center}
\caption{
Error rates for model choice by AIC based on simulation.
Shown are the same distributions of the likelihood ratio statistic $\delta$ as in Figure \ref{fig:pmc}.  
Also shown is
the probability that AIC selects the more complicated model when the simpler is true
(``False Positives'', light shading);
and the probability that AIC selects the more simpler model when the more complicated is true
(``False Negatives'' error, dark shading).
% The dotted line represents the AIC threshold -- any likelihood ratios that exceed this difference favor the more complex model under the AIC criterion.  In each comparison, a significant fraction of the distribution of likelihood ratios observed for data simulated under the simple model (blue curve) falls to the right of this threshold, corresponding to Type I error (orange shading).  The red curve corresponding to simulations of the complex model occasionally generates data that falls to left of the AIC threshold, particularly when the phylogeny is relatively uninformative for the comparison.  This corresponds to the Type II error rate, shaded in yellow.  Error rates for AIC and BIC are shown in Table~\ref{tab:aic_errors}.
}
\label{fig:aic_errors}
\end{figure}

We can compare this method to 
information criteria (e.g.\ AIC, BIC), 
which are the standard tools for model comparison in comparative methods of continuous traits~\citep{Butler2004}.  
Because we have generated simulated datasets under both hypothesized models, 
it is straightforward to estimate how often these datasets
are misclassified by various information criteria. 
The same distributions from Figure~\ref{fig:pmc} are shown 
with the cutoff given by AIC for choosing the more complex model in Figure~\ref{fig:aic_errors}.
We see that AIC would assign nearly half (47.7\%) of the simulations done under OU.3 incorrectly to the OU.15 model,
and that the observed data would also be assigned to OU.15.
If we evaluate the performance of AIC when comparing two reasonable models, OU.3 and OU.4,
information criteria still prefer the more complicated model
(AIC(OU.3) =-39.6; AIC(OU.4) =-41.3, and BIC(OU.3) =-33.9; BIC(OU.4) = -34.6),
but here we know this may be illusory, since
% We don't know OU.4 isn't correct, we just know that we don't have good reason to believe it is better than OU.3
Figure~\ref{fig:aic_errors} shows that AIC falsely assigns 44\% of simulations produced under OU.3 as coming from OU.4.
Sample-size correction of AIC (AICc, not shown) can be similarly misleading.  
See the online appendix for example code to reproduce this figure under each of the different information criteria.

\subsection{Applied to non-nested models}
The next example compares OU.3 to OU.4, where as mentioned above,
the degree of overlap between the distributions of $\delta$ under the two models seen in Figure~\ref{fig:pmc}(b)
shows that we have relatively little power to distinguish between the two.
Note that
since the painting defining the OU.4 model is not a refinement of the painting defining the OU.3 model,
the two models are not nested.
The Monte Carlo approach applies equally well to non-nested models,
unlike the asymptotic derivations commonly used to justify information criteria.
We furthermore do not have to determine the difference in number of parameters, 
as is required by AIC, which in some situations is not obvious.

\begin{table}
\centering
\begin{tabular}{|l|ll|ll|ll|}
\hline
& \multicolumn{2}{|c|}{AIC errors (\%)} & \multicolumn{2}{|c|}{BIC errors (\%)}  & \multicolumn{2}{|c|}{AICc errors (\%)} \\ 
\hline
Comparison & Type I  & Type II & Type I & Type II & Type I & Type II \\
\hline
BM vs.\ OU.3   & 37.00 & 0.00  & 15.90 & 0.45  & 13.05 & 1.05  \\
OU.3 vs.\ OU.4  & 43.75 & 8.25  & 29.35 & 14.5  & 2.30  & 73.55 \\
OU.3 vs.\ OU.15 & 47.75 & 0.00  & 13.65 & 0.00  & 0.00  & 100   \\
BM vs.\ OU.1   & 19.95 & 76.65 & 11.95 & 86.05 & 8.90  & 89.7  \\
\hline
\end{tabular}
\caption{A comparison of error rates across various information criteria. In the comparisons that have high overlap between the distributions (BM vs.\ OU.1, OU.3 vs.\ OU.4, Figure~\ref{fig:pmc}), at least one of the rates will be high for any method.  In cases with adequate power (OU.3 vs.\ OU.15, BM vs.\ OU.3), information criteria can still have high error rates.  The methods we describe allow the researcher not only to estimate these rates, but to specify a tradeoff between the error types.
}
\label{tab:aic_errors}
\end{table}

\begin{figure}[hht]
\begin{center}
    \subfigure[ ]{\includegraphics[width=3in]{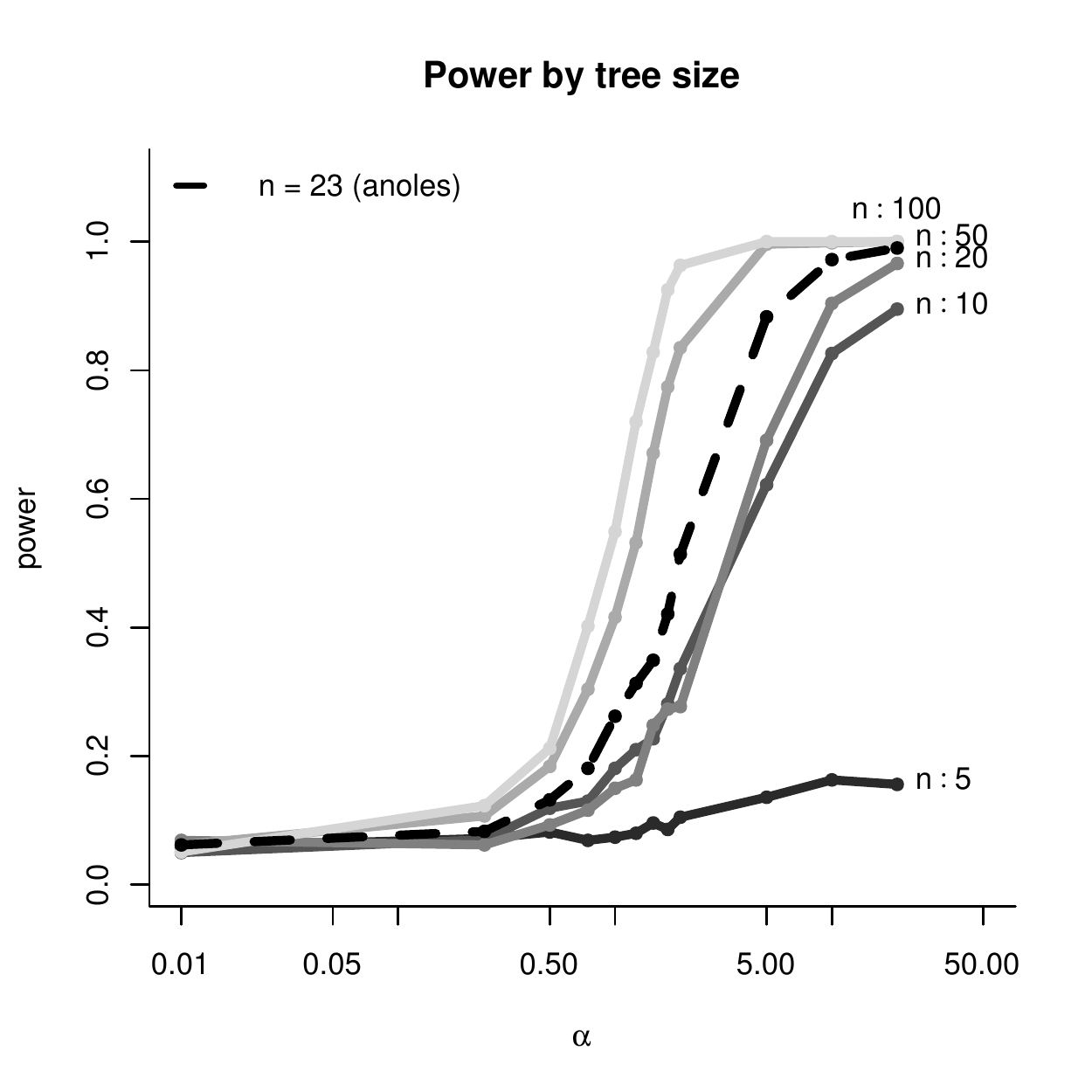} \label{fig:powercurve_size} }
    \subfigure[ ]{\includegraphics[width=3in]{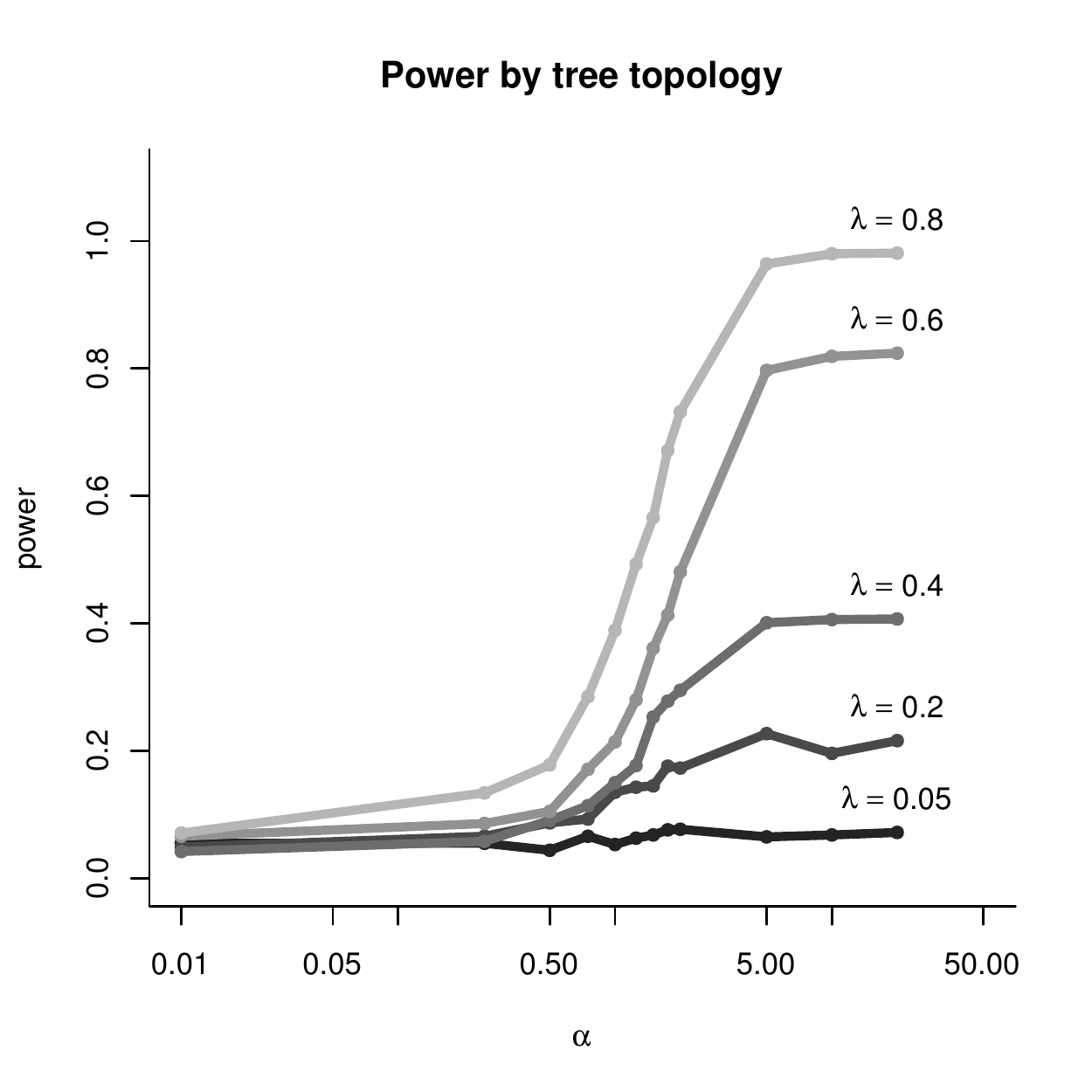} \label{fig:powercurve_shape} }
\end{center}
\caption{
Power to identify stabilizing selection $\alpha$ at a given strength on different phylogenies.  
Shown is the empirical probability that data generated with a given $\alpha$ on a given tree 
will favor a OU.1 model over BM, 
based on a cutoff of the likelihood ratio statistic $\delta$ chosen 
to have a false positive probability of 5\%,
based on 1000 simulations with $\sigma=1$.
(a) Increasing the number of taxa in the tree (simulated under a birth-death model) increases the power to detect a given strength of selection.  
(b) Fixing the number of taxa to 50, we distort the shape of the simulated tree to one in which most of the branching events occur farther and farther in the past using Pagel's $\lambda$ transformation.  On trees that are highly distorted (smaller $\lambda$) we have substantially less power to detect any given strength of selection.  
}
\label{fig:power}
\end{figure}

\subsection{When the data are insufficient to distinguish between models} \label{sect:power}
The fourth comparison is between the simplest models, BM and OU.1. 
Figure~\ref{fig:pmc}(d)
shows that there is essentially no information to adequately distinguish between them.
This should not be taken as evidence that BM is a better fit,
but rather that 
given the small selection parameter estimated from the anoles data,
we have low power to distinguish OU.1 from BM on this phylogeny.  
The strength of selection in the OU model is represented by $\alpha$ in equation \eqref{ou},
and is measured in units of inverse time since the common ancestor (when the tree height has been normalized to unity).
Hence the maximum likelihood estimate for this model with a value of $\alpha = 0.2$
means that correlations between traits that diverged at that common ancestor will have decayed to only $e^{-0.2} =.81$ 
of what is expected under BM.
The chance we could detect this level of selection at 95\% false positive rate (\emph{i.e.} the power) was only 7\%. 

What is the weakest level of stabilizing selection on a trait we could reliably detect using this {\it Anolis} phylogeny?  
To answer this, we repeat the analysis on data simulated using OU.1 models
with progressively larger $\alpha$
and estimate the power for each.
The results are shown as the dashed curve in Figure~\ref{fig:powercurve_size}. 
Power increases with increasing strength of selection $\alpha$,
which we can visualize by imagining the darker distribution of Figure~\ref{fig:pmc}(d) moving farther to the right.
In the next section,
we use this approach of power simulation to understand what aspects of phylogeny 
(\emph{i.e.} shape and size) influence its power to detect a given strength of selection.

\section{Understanding the role of phylogeny shape and size on estimates of selection}
\label{ss:shapesize}

The shape and size of the phylogeny is key to understanding how much information about evolutionary processes 
it is possible to extract from characters of taxa at the leaves of the tree. 
As an application of the method of obtaining a power curve for the strength of selection described in section \ref{sect:power} 
we can compare the power curves for trees of different shapes.
As before, we are comparing
the single--optimum Ornstein-Uhlenbeck (OU.1) model to the Brownian motion (BM) model without selection,
and computing the power to correctly choose the OU.1 model at different values of $\alpha$,
if we choose models based on the 95\% quantile of $\delta$ under the BM model.
% In Figure~\ref{fig:power} we show the power curves for different simulated phylogenies, each normalized to have unit length,.
% For each point in the figure, replicates are simulated from the OU process with the given $\alpha$, then be fit to both OU and BM models and compared to the null distribution of likelihood ratios simulated from BM.
Figure~\ref{fig:powercurve_size} compares trees simulated from 
a pure-birth process with increasing number of taxa, scaled to unit height.
%One conclusion we can draw from the figure is that detecting selection that 
%operates on timescales any shorter than the height of the tree ($\alpha$ values less than unity) requires over 100 taxa.  

Number of taxa is not all that matters;
Figure~\ref{fig:powercurve_shape} considers a single (simulated pure-birth) tree of 50 taxa rescaled 
so that successively more of the time occurs in the tips and so that the speciation events occur more distantly in the past.
The farther in the past diversification has occurred, the less informative the tree.
This is the rescaling performed by the $\lambda$ transformation described in section \ref{ss:models}.
Covariances introduced by different amounts of shared evolution are crucial 
for distinguishing slower character diversification rates $\sigma$ from stronger selection $\alpha$.  
We see that as the branching events occur earlier (smaller $\lambda$ transformations),
these correlations are harder to detect,
so the phylogeny becomes less informative.

% err, still not sure about this paragraph
We note that many simulation studies~\citep[\emph{e.g.}][]{Freckleton2002}
are conducted using trees generated by a pure-birth (Yule) process, which generates phylogenies 
with more very shallow nodes than are generally seen in practice.
Perhaps counter-intuitively, the presence of these highly-correlated points makes the phylogenies particularly informative
relative to branching patterns resulting from any density-dependent or niche-filling models. % conversely, birth-death are more accurate, pull of the present 
Early bursts of speciation such as adaptive radiations 
will tend to generate phylogenies that are less informative of parameters such as the strength of selection, $\alpha$.  
These examples show that the ability to distinguish between models can depend strongly on the value of the parameters, 
the number of taxa, and the shape of the tree.  
Rather than attempt to draw rules of thumb from such exercises, 
we suggest that it is best to perform a power analysis that is specific to the phylogeny and estimated model parameters being compared.  
%\citet{Harmon2010}.  
% Capturing both the influence of number of taxa and structure on the ability to detect selection, this provides a convenient way to visualize the power in a phylogenetic tree.   

% Rates / Alpha harder to estimate than means, variances.  
% Show the parameter uncertainty
% Other tests: symmetric differences in Brownian rate: show that Brownian

\section{Discussion}
We have introduced a general, simulation-based method to choose between models of character evolution and quantify the power of such choices on a particular phylogeny.
While the methodological underpinnings of this approach are not new, the field of comparative methods continues to rely almost universally on information criteria.  We have illustrated that the performance of these methods can be remarkably poor, particularly with insufficiently large or structured phylogenies.  The results can provide a clear indication of when a phylogenetic tree is either too small or too unstructured to resolve differences in the proposed models.  

Though our analysis selects the same model (OU.3) for the anoles dataset as does \citet{Butler2004},
we have shown that existing approaches such as AIC \citet[as used in][]{Butler2004}
would have preferred either of our more complex models (OU.4 and OU.15).
Our models are chosen to illustrate various possible outcomes:
not only can we choose either the simpler or the more complex model, 
but through power simulations we can determine if choice of simpler model 
is due to poor fit of the data by the complex model, 
or simply due to insufficient data.
% These results highlight the importance of performing this power analysis on a case-by-case basis.

Since their introduction in a modeling framework in~\citet{Felsenstein1985}, 
phylogenetic comparative methods have continued to increase in complexity.  
We provide a simple method to reliably indicate if the informativeness of the datasets is keeping pace with this increase in complex models.  
Through these methods, 
we can know when the comparison we are making is too fine for the resolution of the data,
as in the BM vs.\ OU.1 comparison Figure~\ref{fig:pmc}(d), and when increased model complexity is clearly unsupported,
as in OU.3 vs.\ OU.15 comparison, Figure~\ref{fig:pmc}(c). 
Model choice plays a similar role in many other models in comparative phylogenetics, 
such as deciding between the various tree transforms such as $\lambda$, $\delta$, $\gamma$, or ACDC,
which can benefit from the same attention to whether the data are adequately informative. 

As shown in section \ref{ss:shapesize}, the power to distinguish between two models
can depend strongly on the parameter values,
which can be a subtle point and pose difficulties for interpretation.
For instance, 
if a power analysis is done by simulating under a certain set of parameter values,
but the test is applied to datasets consistent with very different parameter values
\citep[a situation found in][]{Harmon2010},
then it remains a possibility that failure to find evidence for more complex models
results from a lack of power.
%\plr{How's this sound? Is it useful?}
%\cdb{I think it still sounds confusing.}

Our results cast doubt on the use of AIC for phylogenetic model selection;
however, mathematically our methods are very similar to information criteria.
When applied to a pair of models, the various information criteria (AIC, BIC, AICc, etc.)
give a cutoff for the likelihood ratio statistic $\delta$
that determines which model to choose.
Our method can provide such a cutoff as well,
but also allows choice of such a cutoff based on the power--false positive tradeoff.
% As seen for instance in \citet{Harmon2010}, 
One use for our methods would be to simply
quantify the resolving power of an AIC-based model choice.
A drawback of our method over AIC is that it does not compare simultaneously many models,
instead relying on a collection of pairwise comparisons.
This is a disadvantage particularly when AIC is applied to find the best model out of many,
and the goal is to find a parsimonious predictive model of more complex reality.
However, it seems to us that comparative methods are usually concerned with 
rigorously distinguishing between alternative models, and so
the goal of model choice is to describe underlying process 
rather than to provide plausible predictions.
See \citet{Burnham2002} for discussion of a philosophy of model selection using AIC in a predictive framework.  
%\cdb{Is this soft-pedaling too much to be useful?  }

The procedure we describe is grounded in a familiar maximum-likelihood framework of model comparison, 
and the dependence on certain estimated parameter values for each model poses one of the difficulties for interpretation.
A Bayesian approach might compare models using Bayes factors, 
thus integrating over all parameter values for each model,
and could be implemented using a reversible jump Markov chain Monte Carlo scheme~\citep{Green1995a}.
Note, however, that the restriction to fixed parameter values is not necessarily a limitation,
as it allows us to
perform such analyses as identifying the weakest level of selection detectable on a given phylogeny,
as in the power curves of Figure~\ref{fig:power}.
% A Bayesian framework brings its own conceptual and computational challenges, 
% such as the choice of appropriate priors and estimates of convergence.
%; this remains a promising direction for future work.  

Comparative data, while an integral and powerful tool in evolutionary biology,
sometimes holds only limited information about the evolutionary process.
We suggest that the application of these approaches to specific dataset should routinely be guided 
by the use of simulation to assess model choice and power.
%This will allow a more tailored approach to guide the use and extension of methods and datasets.

\subsection{A parallelized package for the computational methods}
To compare models using information criteria it is only necessary to fit each model to the observed data once,
while the Monte Carlo approach we describe requires $2n$ model simulations and $4n$ model fits, where $n$ is the number of replicates used.
Fortunately, fitting is both fast and easy to parallelize on modern architectures.
Our R package \texttt{pmc} integrates parallel computation (from the \texttt{snowfall} package) 
with commonly used phylogenetic model fitting tools
provided in the \texttt{geiger}, \texttt{ape} and \texttt{ouch} packages. 
The analyses presented in this paper are included as examples, most of which can be run in minutes when spread over many processors.   

\subsection{Guidelines for analysis}
We have discussed how to compare models pairwise,
and applied the methods to a series of models for the {\it Anolis} dataset.
However, we have not discussed what one is to do when faced with a multitude of models.
Here, as in the situation of choosing which variables to use in a multiple linear regression,
there is no single best answer.
If there are few enough models, by analogy to stepwise addition for linear regression,
one could arrange the models in rough order of complexity, 
begin with the simplest,
and compare each to the next more complex, 
stopping when there is insufficient support to choose a more complex model.
Alternatively, one could do all pairwise comparisons,
although the results may be difficult to interpret if there no single model is clearly best.
If there are many models, 
one option would be to rank all models according to AIC score,
and evaluate uncertainty by comparing each model to the top-ranking few models by our methods.
There are many methods and philosophies of model choice;
it is our opinion that a good method of evaluating uncertainties behind model choice
can only aid in this process.

\section{Acknowledgements}
CB thanks P.\ Wainwright and the rest of the Wainwright lab for valuable input and inspiration 
and the generous support of the Computational Sciences Graduate Fellowship from the Department of Energy under grant number DE-FG02-97ER25308. 
PR was supported by funds from G.\ Coop and S.\ Schreiber, and by a NIH fellowship under grant number F32GM096686.
GC is supported in part by a Sloan Fellowship in Computational and Evolutionary Molecular Biology and by University of California Davis start-up funds. 
We also thank Luke Harmon and an anonymous reviewer for their insightful comments on an earlier version of the manuscript.  
% Should we acknowledge Luke by name or not? should I ask him?
\section*{ }%bibliography
%\bibliography{library.bib}
%\bibliography{library}
%\bibliography{Math\ Methods\ and\ Model\ Inference}

\begin{thebibliography}{34}
\expandafter\ifx\csname natexlab\endcsname\relax\def\natexlab#1{#1}\fi
\expandafter\ifx\csname url\endcsname\relax
  \def\url#1{\texttt{#1}}\fi
\expandafter\ifx\csname urlprefix\endcsname\relax\def\urlprefix{URL }\fi

\bibitem[{Blomberg et~al.(2003)Blomberg, Garland, and Ives}]{Blomberg2003}
Blomberg, S., Garland, J.~T., Ives, A., 2003. {Testing for phylogenetic signal
  in comparative data: behavioral traits are more labile}. Evolution;
  international journal of organic evolution 57~(4), 717--745.
\newline\urlprefix\url{http://www3.interscience.wiley.com/journal/118867878/ab%
stract}

\bibitem[{Burnham and Anderson(2002)}]{Burnham2002}
Burnham, K.~P., Anderson, D., 2002. {Model Selection and Multi-Model
  Inference}. Springer.
\newline\urlprefix\url{http://www.amazon.com/Selection-Multi-Model-Inference-K%
enneth-Burnham/dp/0387953647}

\bibitem[{Butler and King(2004)}]{Butler2004}
Butler, M.~A., King, A.~A., Dec. 2004. {Phylogenetic Comparative Analysis: A
  Modeling Approach for Adaptive Evolution}. The American Naturalist 164~(6),
  683--695.
\newline\urlprefix\url{http://www.jstor.org/stable/10.1086/426002}

\bibitem[{Cox(1961)}]{Cox1961}
Cox, D.~R., 1961. {Tests of Seperate Families of Hypotheses}. In: Proceedings
  of the 4th Berkeley Symposium, University of California Press. No.~2. pp. 105
  -- 123.

\bibitem[{Cox(1962)}]{Cox1962}
Cox, D.~R., 1962. {Further results on tests of separate families of
  hypotheses}. Journal of the Royal Stastical Society 24~(2), 406--424.
\newline\urlprefix\url{http://www.jstor.org/stable/2984232}

\bibitem[{Diciccio and Efron(1996)}]{Diciccio1996}
Diciccio, T.~J., Efron, B., 1996. {Bootstrap Confidence Intervals}. Statistical
  Science 11~(3), 189--212.

\bibitem[{Edwards and Cavalli-Sforza(1964)}]{Edwards1964}
Edwards, A.~W., Cavalli-Sforza, L.~L., 1964. {Reconstruction of evolutionary
  trees}. In: {V. H. Heywood}, McNeill, J. (Eds.), Phenetic and Phylogenetic
  Classification. Systematists Association, London, pp. 67--76.

\bibitem[{Efron(1987)}]{Efron1987}
Efron, B., 1987. {Better bootstrap confidence intervals}. Journal of the
  American Statistical Association 82~(397), 171-- 185.
\newline\urlprefix\url{http://www.jstor.org/stable/2289144}

\bibitem[{Felsenstein(1985)}]{Felsenstein1985}
Felsenstein, J., 1985. {Phylogenies and the comparative method}. The American
  Naturalist 125~(1), 1.
\newline\urlprefix\url{http://www.journals.uchicago.edu/doi/abs/10.1086/284325}

\bibitem[{Freckleton et~al.(2002)Freckleton, Pagel, and
  Harvey}]{Freckleton2002}
Freckleton, R.~P., Pagel, M., Harvey, P.~H., Dec. 2002. {Phylogenetic analysis
  and comparative data: a test and review of evidence.} The American Naturalist
  160~(6), 712--26.
\newline\urlprefix\url{http://www.ncbi.nlm.nih.gov/pubmed/18707460}

\bibitem[{Garland et~al.(1993)Garland, Dickerman, Janis, and
  Jones}]{Garland1993}
Garland, T., Dickerman, a.~W., Janis, C.~M., Jones, J.~a., Sep. 1993.
  {Phylogenetic Analysis of Covariance by Computer Simulation}. Systematic
  Biology 42~(3), 265--292.
\newline\urlprefix\url{http://sysbio.oxfordjournals.org/cgi/doi/10.1093/sysbio%
/42.3.265}

\bibitem[{Gittleman and Kot(1990)}]{Gittleman1990}
Gittleman, J.~L., Kot, M., 1990. {Adaptation : Statistics and a Null Model for
  Estimating Phylogenetic Effects}. Society of Systematic Biologists 39~(3),
  227--241.

\bibitem[{Goldman(1993)}]{Goldman1993}
Goldman, N., Feb. 1993. {Statistical tests of models of DNA substitution}.
  Journal of Molecular Evolution 36~(2), 182--198.
\newline\urlprefix\url{http://www.springerlink.com/index/10.1007/BF00166252}

\bibitem[{Gorman and Kim(1976)}]{Gorman1976}
Gorman, G.~C., Kim, Y.~J., Mar. 1976. {Anolis Lizards of the Eastern Caribbean:
  A Case Study in Evolution. II. Genetic Relationships and Genetic Variation of
  the Bimaculatus Group}. Systematic Zoology 25~(1), 62.
\newline\urlprefix\url{http://sysbio.oxfordjournals.org/cgi/content/abstract/2%
5/1/62}

\bibitem[{Green(1995)}]{Green1995a}
Green, P., 1995. {\{Reversible jump Markov chain Monte Carlo computation and
  Bayesian model determination\}}. Biometrika 82~(4).
\newline\urlprefix\url{http://biomet.oxfordjournals.org/cgi/content/abstract/8%
2/4/711}

\bibitem[{Hall and Wilson(1991)}]{Hall1991a}
Hall, P., Wilson, S., 1991. {Two guidelines for bootstrap hypothesis testing}.
  Biometrics 47~(2), 757--762.
\newline\urlprefix\url{http://www.jstor.org/stable/2532163}

\bibitem[{Hansen(1997)}]{Hansen1997}
Hansen, T.~F., 1997. {Stabilizing selection and the comparative analysis of
  adaptation}. Evolution; international journal of organic evolution 51~(5),
  1341-- 1351.
\newline\urlprefix\url{http://www.jstor.org/stable/2411186}

\bibitem[{Hansen and Martins(1996)}]{Hansen1996}
Hansen, T.~F., Martins, E.~P., 1996. {Translating between microevolutionary
  process and macroevolutionary patterns: the correlation structure of
  interspecific data}. Evolution; international journal of organic evolution
  50, 1404--1417.

\bibitem[{Harmon et~al.(2010)Harmon, Losos, Davies, Gillespie, Gittleman,
  Jennings, Kozak, McPeek, Moreno-Roark, Near, and Others}]{Harmon2010}
Harmon, L.~J., Losos, J.~B., Davies, T., Gillespie, R., Gittleman, J.,
  Jennings, W., Kozak, K., McPeek, M., Moreno-Roark, F., Near, T., Others,
  2010. {Early bursts of body size and shape evolution are rare in comparative
  data}. Evolution; international journal of organic evolution.
\newline\urlprefix\url{http://www3.interscience.wiley.com/journal/123397103/ab%
stract}

\bibitem[{Harmon et~al.(2008)Harmon, Weir, Brock, Glor, and
  Challenger}]{Harmon2008}
Harmon, L.~J., Weir, J.~T., Brock, C.~D., Glor, R.~E., Challenger, W., 2008.
  {Geiger: investigating evolutionary radiations}. Bioinformatics 24~(1),
  129--131.

\bibitem[{Huelsenbeck and Bull(1996)}]{Huelsenbeck1996}
Huelsenbeck, J.~P., Bull, J.~J., Mar. 1996. {A Likelihood Ratio Test to Detect
  Conflicting Phylogenetic Signal}. Systematic Biology 45~(1), 92--98.
\newline\urlprefix\url{http://sysbio.oxfordjournals.org/cgi/content/abstract/4%
5/1/92}

\bibitem[{Lazell(1972)}]{Lazell1972}
Lazell, J.~D., 1972. {The Anoles (Sauria, Iguanidae) of the Lesser Antilles}.
  Bulletin of the Museum of Comparative Zoology.

\bibitem[{Losos(1990)}]{Losos1990}
Losos, J.~B., 1990. {A Phylogenetic Analysis of Character Displacement in
  Caribbean Anolis Lizards}. Evolution; international journal of organic
  evolution 44~(3), 558--569.

\bibitem[{Losos(2011)}]{Losos2011c}
Losos, J.~B., Jun. 2011. {Seeing the forest for the trees: the limitations of
  phylogenies in comparative biology (american society of naturalists address)
  *.} The American naturalist 177~(6), 709--27.
\newline\urlprefix\url{http://www.ncbi.nlm.nih.gov/pubmed/21597249}

\bibitem[{McLachlan(1987)}]{McLachlan1987}
McLachlan, G.~J., 1987. {On Bootstrapping the Likelihood Ratio Test Stastistic
  for the Number of Components in a Normal Mixture}. Applied Statistics 36~(3),
  318.
\newline\urlprefix\url{http://www.jstor.org/stable/2347790}

\bibitem[{Neyman and Pearson(1933)}]{Neyman1933}
Neyman, J., Pearson, E., 1933. {On the problem of the most efficient tests of
  statistical hypotheses}. Philosophical Transactions of the Royal Society of
  London. Series A, Containing Papers of a Mathematical or Physical Character
  231~(694-706), 289.
\newline\urlprefix\url{http://rsta.royalsocietypublishing.org/content/231/694-%
706/289.full.pdf}

\bibitem[{O'Meara et~al.(2006)O'Meara, An\'{e}, Sanderson, and
  Wainwright}]{O'Meara2006}
O'Meara, B.~C., An\'{e}, C., Sanderson, M.~J., Wainwright, P.~C., May 2006.
  {Testing for different rates of continuous trait evolution using likelihood.}
  Evolution; international journal of organic evolution 60~(5), 922--33.
\newline\urlprefix\url{http://www.ncbi.nlm.nih.gov/pubmed/16817533}

\bibitem[{Pagel(1994)}]{Pagel1994}
Pagel, M., 1994. {Detecting Correlated Evolution on Phylogenies: A General
  Method for the Comparative Analysis of Discrete Characters}. Proceedings of
  The Royal Society B 255, 37--45.

\bibitem[{Pagel(1999)}]{Pagel1999}
Pagel, M., Oct. 1999. {Inferring the historical patterns of biological
  evolution.} Nature 401~(6756), 877--84.
\newline\urlprefix\url{http://www.ncbi.nlm.nih.gov/pubmed/10553904}

\bibitem[{Price(1997)}]{Price1997}
Price, T., 1997. {Correlated evolution and independent contrasts.} Proceedings
  of The Royal Society B 352~(1352), 519--529.
\newline\urlprefix\url{http://www.pubmedcentral.nih.gov/articlerender.fcgi?art%
id=1691942}

\bibitem[{Revell(2010)}]{Revell2010}
Revell, L.~J., Jun. 2010. {Phylogenetic signal and linear regression on species
  data}. Methods in Ecology and Evolution (online-fi, no--no.
\newline\urlprefix\url{http://blackwell-synergy.com/doi/abs/10.1111/j.2041-210%
X.2010.00044.x}

\bibitem[{Revell and Harmon(2008)}]{Revell2008a}
Revell, L.~J., Harmon, L.~J., 2008. {Testing quantitative genetic hypotheses
  about the evolutionary rate matrix for continuous characters}. Evolutionary
  Ecology Research 10~(3), 311--331.
\newline\urlprefix\url{http://anolis.oeb.harvard.edu/~liam/pdfs/Revell\_and\_H%
armon\_2008.EER.pdf}

\bibitem[{Schneider et~al.(2001)Schneider, Losos, Queiroz, Journal, Mar, and
  Queiroz}]{Schneider2001}
Schneider, C.~J., Losos, J.~B., Queiroz, K.~D., Journal, S., Mar, N., Queiroz,
  K. D.~E., 2001. {Evolutionary Relationships of the Anolis bimaculatus Group
  from the Northern Lesser Antilles}. Journal of Herpetology 35~(1), 1--12.

\bibitem[{Stenson et~al.(2004)Stenson, Thorpe, and Malhotra}]{Stenson2004}
Stenson, A.~G., Thorpe, R.~S., Malhotra, A., Jul. 2004. {Evolutionary
  differentiation of bimaculatus group anoles based on analyses of mtDNA and
  microsatellite data.} Molecular phylogenetics and evolution 32~(1), 1--10.
\newline\urlprefix\url{http://www.ncbi.nlm.nih.gov/pubmed/15186792}

\end{thebibliography}

\end{document}